\documentclass{rspublic}
\begin{document}

\input epsf

\title[The Origin of Matter and Structure in the Universe]{The 
Origin of Matter and Structure in the Universe}

\author[Juan Garc\'\i a-Bellido]{Juan Garc\'\i a-Bellido}

\affiliation{Theoretical Physics Group, Blackett Laboratory,
Prince Consort Road, \\ Imperial College, London SW7 2BZ, UK \\
e-mail: bellido@ic.ac.uk}

\label{firstpage}

\maketitle

\begin{abstract}{early universe; inflation; cosmic microwave background;
cosmological parameters; large scale structure; gravitational waves;
baryogenesis; dark matter} 

Cosmology is nowadays one of the most active areas of research in
fundamental science. We are going through a true revolution in the
observations that are capable of providing crucial information about the
origin and evolution of the universe. In the first years of the next
millenium we will have, for the first time in the history of such an
ancient science as cosmology, a precise knowledge about a handful of
parameters that determine our Standard Cosmological Model. This standard
model is based on the inflationary paradigm, a period of exponential
expansion in the early universe responsible for the large scale
homogeneity and flatness of our observable patch of the universe. A
spectrum of density perturbations, seen in the microwave background as
temperature anisotropies, could have been produced during inflation from
quantum fluctuations that were stretched to cosmological size by the
expansion, and later gave rise, via gravitational collapse, to the
observed large scale structure of clusters and superclusters of
galaxies. Furthermore, the same theory predicts that all the matter and
radiation in the universe today originated at the end of inflation from
an explosive production of particles that could also have been the
origin of the present baryon asymmetry, before the universe reached
thermal equilibrium at a very large temperature. From there on, the
universe cooled down as it expanded, in the way described by the
standard hot big bang model. With the observations that will soon become
available in the next millenium, we will be able to test the validity of
the inflationary paradigm, and determine with unprecedented accuracy the
parameters of a truly Standard Model of Cosmology.

\end{abstract}

\section{Introduction}

Our present understanding of the universe is based upon the successful
hot big bang theory, which explains its evolution from the first
fraction of a second to our present age, around 13 billion years
later. This theory rests upon four strong pillars, a theoretical
framework based on general relativity, as put forward by Albert Einstein
and Alexander A. Friedmann in the 1920s, and three strong observational
facts. First, the expansion of the universe, discovered by Edwin
P. Hubble in the 1930s, as a recession of galaxies at a speed
proportional to their distance to us. Second, the relative abundance of
light elements, explained by George Gamow in the 1940s, mainly that of
helium, deuterium and lithium [See Fig.~1], which were cooked from the
nuclear reactions that took place at around a second to a few minutes
after the big bang, when the universe was a hundred times hotter than
the core of the sun. Third, the cosmic microwave background (CMB), the
afterglow of the big bang, discovered in 1965 by Arno A. Penzias and
Robert W. Wilson as a very isotropic blackbody radiation at a
temperature of about 3 degrees Kelvin (degrees Centigrade above
absolute zero), emitted when the universe was cold enough to form
neutral atoms, and photons decoupled from matter, approximately 300\,000
years after the big bang. Today, these observations are confirmed to
within a few percent accuracy, and have helped establish the hot big
bang as the preferred model of the universe.

The big bang theory could not explain, however, the origin of matter and
structure in the universe; that is, the origin of the matter-antimatter
asymmetry, without which the universe today would be filled by a uniform
radiation continuosly expanding and cooling, with no traces of matter,
and thus without the possibility to form gravitationally bound systems
like galaxies, stars and planets that could sustain life. Moreover, the
standard big bang theory assumes but cannot explain the origin of the
extraordinary smoothness and flatness of the universe on the very large
scales, seen by the microwave background probes and the largest galaxy
catalogs [See Fig.~2]. It cannot explain the origin of the primordial
density perturbations that gave rise to cosmic structures like galaxies,
clusters and superclusters, via gravitational collapse; neither the
quantity and nature of the dark matter that we believe holds the
universe together; nor the origin of the big bang itself.

In the 1980s a new paradigm, deeply rooted in fundamental physics, was
put forward by Alan H. Guth, Andrei D. Linde and others, to address
these fundamental questions. According to the inflationary paradigm, the
early universe went through a period of exponential expansion, driven by
the approximately constant energy density of a scalar field called the
inflaton. In modern physics, elementary particles are represented by
quantum fields, which resemble the familiar electric, magnetic and
gravitational fields. A field is simply a function of space and time
whose quantum oscillations are interpreted as particles. For instance,
the photon is the particle associated with the electromagnetic field. In
our case, the inflaton field has associated with it a large potential
energy density, which drives the exponential expansion during inflation
[See Fig.~3]. We know from general relativity that the density of matter
determines the expansion of the universe, but a constant energy density
acts in a very peculiar way: as a repulsive force that makes any two
points in space separate at exponentially large speeds. (This does not
violate the laws of causality because there is no information carried
along in the expansion, it is simply the stretching of spacetime.)

This superluminal expansion is capable of explaining the large scale
homogeneity of our observable universe and, in particular, why the
microwave background looks so isotropic: regions separated today by more
than a degree in the sky were, in fact, in causal contact before
inflation, but were stretched to cosmological distances by the expansion
[See Fig.~4]. Any inhomogeneities present before the tremendous
expansion would be washed out. Moreover, in the usual big bang scenario
a flat universe, one in which the gravitational attraction of matter is
exactly balanced by the cosmic expansion, is unstable under
perturbations: a small deviation from flatness is amplified and soon
produces either an empty universe or a collapsed one. For the universe
to be nearly flat today, it must have been extremely flat at
nucleosynthesis for example, deviations not exceeding more than one
part in $10^{15}$. This extreme fine tuning of initial conditions was
also solved by the inflationary paradigm [See Fig.~5]. Thus inflation is an
extremely elegant hypothesis that explains how a region much, much
greater that our own observable universe could have become smooth and
flat without recourse to {\em ad hoc} initial conditions.

\section{The origin of structure in the universe}

If cosmological inflation made the universe so extremely flat and
homogeneous, where did the galaxies and clusters of galaxies come from?
One of the most astonishing predictions of inflation, one that was not
even expected, is that quantum fluctuations of the inflaton field are
stretched by the exponential expansion and generate large scale
perturbations in the metric. Inflaton fluctuations are small
wave-packets of energy that, according to general relativity, modify the
spacetime fabric, creating a whole spectrum of curvature
perturbations. The use of the word spectrum here is closely related to
the case of light waves propagating in a medium: a spectrum
characterises the amplitude of each given wavelength. In the case of
inflation, the inflaton fluctuations induce waves in the spacetime
metric that can be decomposed into different wavelengths, all with
approximately the same amplitude, that is, corresponding to a
scale-invariant spectrum. These patterns of perturbations in the metric
are like fingerprints that characterise unequivocally a period of
inflation. When matter fell in the troughs of these waves, it created
density perturbations that collapsed gravitationally to form galaxies,
clusters and superclusters of galaxies, with a spectrum that is also
scale-invariant. Such a type of spectrum was proposed in the early 1970s
(before inflation) by Edward R. Harrison, and independently by the
Russian cosmologist Yakov B. Zel'dovich, to explain the distribution of
galaxies and clusters of galaxies on very large scales in our observable
universe.

Various telescopes, like the Hubble Space Telescope, the twin Keck
telescopes in Hawaii and the European Southern Observatory telescopes in
Chile, are exploring the most distant regions of the universe and
discovering the first galaxies at large distances.  According to the big
bang theory, the further the galaxy is, the larger its recession
velocity, and the larger the shift towards the red of the spectrum of
light from that galaxy. Astronomers thus measure distances in units of
red-shift $z$. The furthest galaxies observed so far are at redshifts of
$z\simeq5$, or 12 billion light-years from the Earth, whose light was
emitted when the universe had only about 5 percent of its present age.
Only a few galaxies are known at those redshifts, but there are at
present various catalogs like the CfA and APM galaxy catalogs, and more
recently the IRAS PSCz [See Fig.~2] and Las Campanas redshift surveys,
that study the spatial distribution of hundreds of thousands of galaxies
up to distances of a billion light-years, or $z<0.1$, that
receed from us at speeds of tens of thousands of kilometres per
second. These catalogs are telling us about the evolution of clusters of
galaxies in the universe, and already put constraints on the theory of
structure formation based on the gravitational collapse of the small
inhomogeneities produced during inflation. From these observations one
can infer that most galaxies formed at redshifts of order 2 to 4;
clusters of galaxies formed at redshifts of order one, and superclusters
are forming now. That is, cosmic structure formed from the bottom up,
from galaxies to clusters to superclusters, and not the other way
around.

This fundamental difference is an indication of the type of matter that
gave rise to structure. We know from primordial nucleosynthesis that all
the baryons in the universe cannot account for the observed amount of
matter, so there must be some extra matter (dark since we don't see it)
to account for its gravitational pull. Whether it is relativistic (hot)
or non-relativistic (cold) could be inferred from observations:
relativistic particles tend to diffuse from one concentration of matter
to another, thus transferring energy among them and preventing the
growth of structure on small scales. This is excluded by observations,
so we conclude that most of the matter responsible for structure
formation must be cold. How much there is is a matter of debate at the
moment. Some recent analyses suggest that there is not enough cold dark
matter to reach the critical density required to make the universe
flat. Some other form of energy permeates the universe, if we want to
make sense of the present observations. In order to resolve this issue,
even deeper galaxy redshift catalogs are underway, looking at millions
of galaxies, like the Sloan Digital Sky Survey (SDSS) and the
Anglo-Australian two degree field (2dF) Galaxy Redshift Survey, which
are at this moment taking data, up to redshifts of $z<0.5$, or several
billion light-years away, over a large region of the sky. These
important observations will help astronomers determine the nature of the
dark matter and test the validity of the models of structure formation.

\begin{table}\label{table1}
\caption{The parameters of the standard cosmological model}
\longcaption{The standard model of cosmology has around 12 different
parameters, needed to describe the background spacetime, the matter
content and the spectrum of density perturbations. We include here the
present range of the most relevant parameters, and the percentage error
with which the microwave background probes MAP and Planck (without
polarisation) will be able to determine them in the near future.
The rate of expansion is written in units of $H=100\,h$ km/s/Mpc}
\begin{tabular}{lllll}
\hline
physical quantity & 
\multicolumn{1}{c}{symbol} &
\multicolumn{1}{c}{present range} &
\multicolumn{1}{c}{MAP} &
\multicolumn{1}{c}{Planck} \\
\hline
\hline
luminous matter & 
\multicolumn{1}{c}{$\Omega_{\rm lum}h^2$} &
\multicolumn{1}{c}{$0.001 - 0.005$} &
\multicolumn{1}{c}{$-$} &
\multicolumn{1}{c}{$-$} \\
\hline
baryonic matter & 
\multicolumn{1}{c}{$\Omega_{\rm B}h^2$} &
\multicolumn{1}{c}{$0.01 - 0.03$} &
\multicolumn{1}{c}{5\%} &
\multicolumn{1}{c}{0.6\%} \\
\hline
cold dark matter & 
\multicolumn{1}{c}{$\Omega_{\rm M}h^2$} &
\multicolumn{1}{c}{$0.2 - 1.0$} &
\multicolumn{1}{c}{10\%} &
\multicolumn{1}{c}{0.6\%} \\
\hline
hot dark matter & 
\multicolumn{1}{c}{$\Omega_\nu h^2$} &
\multicolumn{1}{c}{$0 - 0.3$} &
\multicolumn{1}{c}{5\%} &
\multicolumn{1}{c}{2\%} \\
\hline
cosmological constant & 
\multicolumn{1}{c}{$\Omega_\Lambda h^2$} &
\multicolumn{1}{c}{$0 - 0.8$} &
\multicolumn{1}{c}{8\%} &
\multicolumn{1}{c}{0.5\%} \\
\hline
spatial curvature & 
\multicolumn{1}{c}{$\Omega_0 h^2$} &
\multicolumn{1}{c}{$0.2 - 1.5$} &
\multicolumn{1}{c}{4\%} &
\multicolumn{1}{c}{0.7\%} \\
\hline
rate of expansion & 
\multicolumn{1}{c}{$h$} &
\multicolumn{1}{c}{$0.4 - 0.8$} &
\multicolumn{1}{c}{11\%} &
\multicolumn{1}{c}{2\%} \\
\hline
age of the universe & 
\multicolumn{1}{c}{$t_0$} &
\multicolumn{1}{c}{$11 - 17$ Gyr} &
\multicolumn{1}{c}{10\%} &
\multicolumn{1}{c}{2\%} \\
\hline
spectral amplitude & 
\multicolumn{1}{c}{$Q_{\rm rms}$} &
\multicolumn{1}{c}{$20 - 30\ \mu$K } &
\multicolumn{1}{c}{0.5\%} &
\multicolumn{1}{c}{0.1\%} \\
\hline
spectral tilt & 
\multicolumn{1}{c}{$n_{_{\rm S}}$} &
\multicolumn{1}{c}{$0.5 - 1.5$} &
\multicolumn{1}{c}{3\%} &
\multicolumn{1}{c}{0.5\%} \\
\hline
tensor-scalar ratio & 
\multicolumn{1}{c}{$r_{\rm ts}$} &
\multicolumn{1}{c}{$0 - 1.0$} &
\multicolumn{1}{c}{25\%} &
\multicolumn{1}{c}{10\%} \\
\hline
reionisation & 
\multicolumn{1}{c}{$\tau$} &
\multicolumn{1}{c}{$0.01 - 1.0$} &
\multicolumn{1}{c}{20\%} &
\multicolumn{1}{c}{15\%} \\
\hline

\end{tabular}
\end{table}

However, if galaxies indeed formed from gravitational collapse of
density perturbations produced during inflation, one should also expect
to see such ripples in the metric as temperature anisotropies in the
cosmic microwave background, that is, minute deviations in the
temperature of the blackbody spectrum when we look at different
directions in the sky. Such anisotropies had been looked for ever since
Penzias and Wilson's discovery of the CMB, but had eluded all detection,
until NASA's Cosmic Background Explorer (COBE) satellite discovered them
in 1992. The reason why it took so long to discover was that they appear
as perturbations in temperature of only one part in 100\,000. There is,
in fact, a dipolar anisotropy of one part in 1000, in the direction of
the Virgo cluster, but that is interpreted consistently as our relative
motion with respect to the microwave background due to the local
distribution of mass, which attracts us gravitationally towards the
Virgo cluster. When subtracted, we are left with a whole spectrum of
anisotropies in the higher multipoles (quadrupole, octupole, etc.)  
[See Fig.~6]. Soon after COBE, other groups quickly confirmed the
detection of temperature anisotropies at around 30\,$\mu$K, at higher
multipole numbers or smaller angular scales.

There are at this moment dozens of ground and balloon-borne experiments
analysing the anisotropies in the microwave background with angular
resolutions from 10 degrees to a few arc-minutes in the sky. The physics
of the CMB anisotropies is relatively simple: photons scatter off
charged particles (protons and electrons), and carry energy, so they
feel the gravitational potential associated with the perturbations
imprinted in the metric during inflation. An overdensity of baryons
(protons and neutrons) does not collapse under the effect of gravity
until it enters the causal Hubble radius. The perturbation continues to
grow until radiation pressure opposes gravity and sets up acoustic
oscillations in the plasma, very similar to sound waves. Since
overdensities of the same size will enter the Hubble radius at the same
time, they will oscillate in phase. Moreover, since photons scatter off
these baryons, the acoustic oscillations occur also in the photon field
and induces a pattern of peaks in the temperature anisotropies in the
sky, at different angular scales [See Fig.~7]. The larger the amount of
baryons, the higher the peaks. The first peak in the photon distribution
corresponds to overdensities that have undergone half an oscillation,
that is, a compression, and appear at a scale associated with the size
of the horizon at last scattering (when the photons decoupled) or about
one degree in the sky. Other peaks occur at harmonics of this,
corresponding to smaller angular scales. Since the amplitude and
position of the primary and secondary peaks are directly determined by
the sound speed (and hence the equation of state) and by the geometry
and expansion of the universe, they can be used as a powerful test of
the density of baryons and dark matter, and other cosmological
parameters.

By looking at these patterns in the anisotropies of the microwave
background, cosmologists can determine not only the cosmological
parameters, but also the primordial spectrum of density perturbations
produced during inflation. It turns out that the observed temperature
anisotropies are compatible with a scale-invariant spectrum, as
predicted by inflation. This is remarkable, and gives very strong
support to the idea that inflation may indeed be responsible for both
the CMB anisotropies and the large scale structure of the universe.
Different models of inflation have different specific predictions for
the fine details associated with the spectrum generated during
inflation. It is these minute differences that will allow cosmologists
to differentiate among alternative models of inflation and discard those
that do not agree with observations. But most importantly, perhaps, is
that the pattern of anisotropies predicted by inflation is completely
different from those predicted by alternative models of structure
formation, like cosmic defects: strings, vortices, textures, etc. These
are complicated networks of energy density concentrations left over from
an early universe phase transition, analogous to the defects formed in
the laboratory in certain kinds of liquid crystals when they go through
a phase transition. The cosmological defects have spectral properties
very different from those generated by inflation. That is why it is so
important to launch more sensitive, and with better angular resolution,
instruments to determine the properties of the CMB anisotropies.

In the first years of the next millenium two new satellites, the
Microwave Anisotropy Probe (MAP), to be launched by NASA in the year
2000, and Planck Surveyor, due in 2007, by the European Space Agency,
will measure those temperature anisotropies with 100 times better
angular resolution and 10 times better sensitivity than COBE, and thus
allow cosmologists to determine the parameters of the Standard
Cosmological Model with unprecedented accuracy. What makes the microwave
background observations particularly powerful is the absence of large
systematic errors that plague other cosmological measurements. As we
have discussed above, the physics of the microwave background is
relatively simple, compared to, say, the physics of supernova
explosions, and computations can be done consistently within
perturbation theory. Thus most of the systematic errors are theoretical
in nature, due to our ignorance about the primordial spectrum of metric
perturbations from inflation. There is a great effort at the moment in
trying to cover a large region in the parameter space of models of
inflation, to ensure that we have considered all possible alternatives,
like isocurvature or pressure perturbations, non scale invariant or
tilted spectra and non-Gaussian density perturbations. 

In particular, inflation also predicts a spectrum of gravitational
waves. Their amplitude is directly proportional to the total energy
density during inflation, and thus its detection would immediately tell
us about the energy scale (and therefore the epoch in the early
universe) at which inflation occurred. If the period of inflation
responsible for the observed CMB anisotropies is associated with the
Gran Unification scale, when the strong and electroweak interactions are
supposed to unify, then there is a chance that we might see the effect
of gravitational waves in the future satellite measurements, specially
from the analysis of photon polarisation in the microwave background
maps.

Moreover, the stochastic background of gravitational waves generated
during inflation could eventually be observed by the ground-based laser
interferometers like LIGO, VIRGO, GEO, TAMA, etc., which will start
taking data as gravitational wave observatories in the first years of
the next millenium. These are extremely sensitive devices that could
distinguish minute spatial variations, of one part in $10^{23}$ or
better, induced when a gravitational wave from a distant source passes
through the Earth and distorts the spacetime metric.  Gravitational
waves moving at the speed of light are a fundamental prediction of
general relativity. Their existence was indirectly confirmed by Russell
A. Hulse and Joseph H. Taylor, through the precise observations of the
decay in the orbital period of the pulsar PSR1913+16, due to the
emission of gravitational radiation. In the near future, observations of
gravitational waves with laser interferometers will open a completely
new window into the universe. It will allow us to observe with a very
different probe (that of the gravitational interaction) a huge range of
phenomena, from the most violent processes in our galaxy and beyond,
like supernova explosions, neutron star collisions, quasars, gamma ray
bursts, etc., to the origin of the universe.

In our quest for the parameters of the standard cosmological model,
various groups are searching for distant astrophysical objects that can
serve as standard candles to determine the distance to the object from
their observed apparent luminosity. A candidate that has recently been
exploited with great success is a certain type of supernova explosions
at large redshifts. These are stars at the end of their life cycle that
become unstable and violently explode in a natural thermonuclear
explosion that out-shines their progenitor galaxy. The intensity of the
distant flash varies in time, it takes about three weeks to reach its
maximum brightness and then it declines over a period of months.
Although the maximum luminosity varies from one supernova to another,
depending on their original mass, their environment, etc., there is a
pattern: brighter explosions last longer than fainter ones. By studying
the light curves of a reasonably large statistical sample, cosmologists
from two competing groups, the Supernova Cosmology Project and the
High-z Supernova Project, are confident that they can use this type of
supernovae as standard candles. Since the light coming from some of
these rare explosions has travelled for a large fraction of the size of
the universe, one expects to be able to infer from their distribution
the spatial curvature and the rate of expansion of the universe [See
Fig.~8]. One of the surprises that these observations have revealed is
that the universe appears to be accelerating, instead of decelerating as
was expected from the general attraction of matter; something seems to
be acting as a repulsive force on very large scales. The most natural
explanation for this is the existence of a cosmological constant, a
diffuse vacuum energy that permeates all space and, as explained above,
gives the universe an acceleration that tends to separate
gravitationally bound systems from eachother. The origin of such a
vacuum energy is one of the biggests problems of modern physics. Its
observed value is 120 orders of magnitude smaller than predicted by
fundamental high energy physics. If confirmed, it will pose a real
challenge to theoretical physics, one that may affect its most basic
foundations. However, it is still premature to conclude that this is
indeed the case, because of possibly large systematic errors inherent to
most cosmological measurements, given our impossibility to do
experiments under similar circumstances in the laboratory.

\section{The origin of matter in the universe}

Cosmological inflation may be responsible for the metric perturbations
that later gave rise to the large scale structures we see in the
universe, but where did all the matter in the universe come from? why
isn't all in photons which would have inevitably redshifted away in a
cold universe devoid of life? How did we end up being matter dominated?
Everything we see in the universe, from planets and stars, to galaxies
and clusters of galaxies, is made out of matter, so where did the
antimatter in the universe go? Is this the result of an accident, a
happy chance occurrence during the evolution of the universe, or is it
an inevitable consequence of some asymmetry in the laws of nature?
Theorists believe that the excess of matter over antimatter comes from
fundamental differences in their interactions soon after the end of
inflation.

Inflation is an extremely efficient mechanism in diluting any parti\-cle
species or fluctuations. At the end of inflation, the universe is empty
and extremely cold, dominated by the homogeneous coherent mode of the
inflaton. Its potential energy density is converted into particles, as
the inflaton field oscillates coherently around the minimum of its
potential [See Fig.~3]. These particles are initially very far from
equilibrium, but they strongly interact among themselves and soon reach
thermal equilibrium at a very large temperature. From there on, the
universe expanded isoentropically, cooling down as it expanded, in the
way described by the standard hot big bang model. Thus the origin of the
big bang itself, and the matter and energy we observe in the universe
today, can be traced back to the epoch in which the inflaton energy
density decayed into particles. Such a process is called reheating of
the universe. 

Recent developments in the theory of reheating suggest that the decay of
the inflaton energy could be explosive, due to the coherent oscillations
of the inflaton, which induces its stimulated decay. The result is a
resonant production of particles in just a few inflaton oscillations, an
effect very similar to the stimulated emission of a laser beam of
photons. The number of particles produced this way is exponentially
large, which may explain the extraordinarily large entropy, of order
$10^{89}$ particles, in our observable patch of the universe today.
However, the inflaton is supposed to be a neutral scalar field, and thus
its interactions cannot differentiate between particles and
antiparticles. How did we end up with more matter than antimatter? The
study of this cosmological asymmetry goes by the name of baryogenesis
since baryons (mainly protons and neutrons) are the fundamental
constituents of matter in planets, stars and galaxies in the universe
today. So, what are the conditions for baryogenesis?

\begin{table}\label{table2}
\caption{The standard model of particle physics} 
\longcaption{The primary constituents of matter, quarks and leptons, are
divided into three generations. The first generation contains the up and
down quarks and antiquarks, as well as the electron, its neutrino and
their antiparticles.  Ordinary matter is made out almost exclusively of
particles of the first generation: an atom's nucleus contains protons
and neutrons, themselves made of up and down quarks. The other
generations, equally abundant in the early universe, may still exist in
hot environments such as neutron star cores, and are routinely produced
in accelerators.}
\begin{tabular}{lllll}
\hline
& & Transmitters of force & & \\
\hline
\hline
\multicolumn{1}{c}{Weak bosons} &
\multicolumn{1}{c}{Photon} &
\multicolumn{1}{c}{Gluon} &
\multicolumn{1}{c}{Higgs} & \\
\hline
\multicolumn{1}{c}{$W^\pm$\, \ $Z^0$} &
\multicolumn{1}{c}{$\gamma$} &
\multicolumn{1}{c}{$g$} &
\multicolumn{1}{c}{$H^0$} & \\
\hline
\hline
& & Constituents of matter & & \\
\hline
\hline
\multicolumn{1}{c}{Particle} &
\multicolumn{1}{c}{Symbol} &
\multicolumn{1}{c}{Charge} &
\multicolumn{1}{c}{Mass (GeV/$c^2$)} & \\
\hline
First generation & & & &\\
\hline
\multicolumn{1}{c}{up} &
\multicolumn{1}{c}{u} &
\multicolumn{1}{c}{$+2/3$} &
\multicolumn{1}{c}{0.03} & quarks\\
\multicolumn{1}{c}{down} &
\multicolumn{1}{c}{d} &
\multicolumn{1}{c}{$-1/3$} &
\multicolumn{1}{c}{0.06} & \\
\hline
\multicolumn{1}{c}{electron} &
\multicolumn{1}{c}{e} &
\multicolumn{1}{c}{$-1$} &
\multicolumn{1}{c}{0.0005} & leptons\\
\multicolumn{1}{c}{electron neutrino} &
\multicolumn{1}{c}{$\nu_e$} &
\multicolumn{1}{c}{0} &
\multicolumn{1}{c}{0?} & \\
\hline
Second generation & & & & \\
\hline
\multicolumn{1}{c}{charm} &
\multicolumn{1}{c}{c} &
\multicolumn{1}{c}{$+2/3$} &
\multicolumn{1}{c}{1.3} & quarks\\
\multicolumn{1}{c}{strange} &
\multicolumn{1}{c}{s} &
\multicolumn{1}{c}{$-1/3$} &
\multicolumn{1}{c}{0.14} & \\
\hline
\multicolumn{1}{c}{muon} &
\multicolumn{1}{c}{$\mu$} &
\multicolumn{1}{c}{$-1$} &
\multicolumn{1}{c}{0.106} & leptons\\
\multicolumn{1}{c}{muon neutrino} &
\multicolumn{1}{c}{$\nu_\mu$} &
\multicolumn{1}{c}{0} &
\multicolumn{1}{c}{0?} & \\
\hline
Third generation & & & & \\
\hline
\multicolumn{1}{c}{top} &
\multicolumn{1}{c}{t} &
\multicolumn{1}{c}{$+2/3$} &
\multicolumn{1}{c}{174} & quarks\\
\multicolumn{1}{c}{bottom} &
\multicolumn{1}{c}{b} &
\multicolumn{1}{c}{$-1/3$} &
\multicolumn{1}{c}{4.3} & \\
\hline
\multicolumn{1}{c}{tau} &
\multicolumn{1}{c}{$\tau$} &
\multicolumn{1}{c}{$-1$} &
\multicolumn{1}{c}{1.7} & leptons\\
\multicolumn{1}{c}{tau neutrino} &
\multicolumn{1}{c}{$\nu_\tau$} &
\multicolumn{1}{c}{0} &
\multicolumn{1}{c}{0?} & \\
\hline

\end{tabular}
\end{table}

Everything we know about the properties of elementary particles is
included in the standard model of particle physics. It describes more
than a hundred observed particles and their interactions in terms of a
few fundamental constituents: six quarks and six leptons, and their
antiparticles [See Table~2]. The standard model describes three types of
interactions: the electromagnetic force, the strong and the weak nuclear
forces. These forces are transmitted by the corresponding particles: the
photon, the gluon and the W and Z bosons. The theory also requires a
scalar particle, the Higgs particle, responsible for the masses of
quarks and leptons and the breaking of the electroweak symmetry at an
energy scale 100 times the mass of the proton. The Higgs is believed to
lie behind most of the mysteries of the standard model, including the
asymmetry between matter and antimatter.

Symmetries are fundamental properties of any physical theory. A theory
is symmetric under certain symmetry operation, like reflection, if its
laws apply equally well after such an operation is performed on part of
the physical system. An important example is the operation called parity
reversal, denoted by P. It produces a mirror reflection of an object and
rotates it 180 degrees about an axis perpendicular to the mirror. A
theory has P symmetry if the laws of physics are the same in the real
and the parity-reversed world. Particles such as leptons and quarks can
be classified as right- or left-handed depending on the sense of their
internal rotation, or spin, around their direction of motion. If P
symmetry holds, right-handed particles behave exactly like left-handed
ones. The laws of electrodynamics and the strong interactions are the
same in a parity-reflected universe. But, as Chien-Shiung Wu discovered
in 1957, the weak interaction acts very differently on particles with
different handedness: only left-handed particles can decay by means of
the weak interaction, not right-handed ones. Moreover, as far as we
know, there are no right-handed neutrinos, only left-handed. So the weak
force violates P.

Another basic symmetry of nature is charge conjugation, denoted by C.
This operation changes the quantum numbers of every particle into those
of its antiparticle. Charge symmetry is also violated by the weak
interactions: antineutrinos are not left-handed, only right-handed.
Combining C and P one gets the charge-parity symmetry CP, which turns
all particles into their antiparticles and also reverses their
handedness: left-handed neutrinos become right-handed antineutrinos [See
Fig.~9].  Although charge and parity symmetry are individually broken by
the weak interaction, one expects their combination to be
conserved. However, in 1964, a groundbreaking experiment by James
Cronin, Val Finch and Ren\'e Turlay at Brookhaven National Laboratory
showed that CP was in fact violated to one part in 1000. It was hard to
see why CP symmetry should be broken at all and even more difficult to
understand why the breaking was so small. Soon after, in 1972, Makoto
Kobayashi and Toshihide Maskawa showed that CP could be violated within
the standard model if three or more generations of quarks existed,
because of CP nonconserving phases that could not be rotated away. Only
two generations where known at the time, but in 1975 Martin L. Perl and
collaborators discovered the tau lepton at the Stanford Linear
Accelerator Centre (SLAC), the first ingredient of the third generation.
Only recently the last quark in the family was discovered at Fermilab,
the top quark.

But how does this picture fit in the evolution of the universe? In 1967,
the Russian physicist Andrei Sakharov pointed out the three necessary
conditions for the baryon asymmetry of the universe to develop. First,
we need interactions that do not conserve baryon number B, otherwise no
asymmetry could be produced in the first place. Second, C and CP
symmetry must be violated, in order to differentiate between matter and
antimatter, otherwise B nonconserving interactions would produce baryons
and antibaryons at the same rate, thus maintaining zero net baryon
number.  Third, these processes should occur out of thermal equilibrium,
otherwise particles and antiparticles, which have the same mass, would
have equal occupation numbers and would be produced at the same rate.
The standard model is baryon symmetric at the classical level, but
violates B at the quantum level, through the chiral anomaly. Electroweak
interactions violate C and CP, but the magnitude of the latter is
clearly insufficient to account for the observed baryon asymmetry. This
failure suggests that there must be other sources of CP violation in
nature, and thus the standard model of particle physics is probably
incomplete.

One of the most popular extensions of the standard model includes a new
symmetry called supersymmetry, which relates bosons (particles that
mediate interactions) with fermions (the constituents of matter). Those
extensions generically predict other sources of CP violation coming from
new interactions at scales above 1000 times the mass of the proton. Such
scales will soon be explored in the first years of the next millenium by
particle colliders like the Large Hadron Collider (LHC) at CERN, the
European Centre for Particle Physics, and by the Tevatron at Fermilab.
The mechanism for baryon production in the early universe in these
models relies on the strength of the electroweak phase transition, as
the universe cooled and the symmetry was broken. Only for strongly first
order phase transitions is the universe sufficiently far from
equilibrium to produce enough baryon asymmetry. Unfortunately, the phase
transition in these models is invariably too weak to account for the
observed asymmetry, so some other mechanism is needed.

If reheating after inflation occurred in an explosive way, via the
resonant production of particles from the inflaton decay, as recent
developments suggest, then the universe has actually gone through a very
non-linear, non-perturbative and very far from equilibrium stage, before
thermalising via particle interactions. Electroweak baryogenesis could
then take place during that epoch, soon after the end of low energy
inflation at the electroweak scale. Such models can be constructed but
require a specially flat direction (a very small mass for the inflaton)
during inflation, in order to satisfy the constraints from the amplitude
of temperature anisotropies seen by COBE. Such flat directions are
generic in supersymmetric extensions of the standard model. After
inflation, the inflaton acquires a large mass from its interaction with
the Higgs field.

The crucial ingredient of departure from equilibrium, necessary for the
excess production of baryons over antibaryons, is strongly present in
this new scenario of baryogenesis, as the universe develops from a zero
temperature and zero entropy state, at the end of inflation, to a
thermal state with exponentially large numbers of particles, the origin
of the standard hot big bang. If, during this stage, fundamental or
effective interactions that are B, C and CP violating were fast enough
compared to the rate of expansion, the universe could have ended with
the observed baryon asymmetry of one part in $10^{10}$, or one baryon
per $10^9$ photons today, as deduced from observations of the light
element abundances [See Fig.~1]. Recent calculations suggest than indeed,
the required asymmetry could be produced as long as some new physics,
just above the electroweak symmetry breaking scale, induces a new 
effective CP violating interaction.

These new phenomena necessarily involve an interaction between the Higgs
particle, responsible for the electroweak symmetry breaking, and the
inflaton field, responsible for the period of cosmological inflation.
Therefore, for this scenario to work, it is expected that both the Higgs
and the inflaton particles be discovered at the future particle physics
colliders like the LHC and the Next Linear Collider (NLC), to be built
in the next millenium. Furthermore, this new physics would necessarily
involve new interactions in the quark sector, for example inducing CP
violations in the B meson (a bound state composed of a bottom quark and
an antidown quark) system.  Such violations are the main research
objective of the B factory at SLAC in California and at KEK, the High
Energy Accelerator Research Organisation in Tsukuba, Japan. Those
experiments will be taking data in a few months time, at the turn of the
millenium, and could give us clues on the issues of the
matter-antimatter asymmetry and thus on baryogenesis from reheating
after inflation.

If confirmed, such a new scenario of baryogenesis would represent a
leap forward in our understanding of the universe from the unifying
paradigm of inflationary cosmology. Furthermore, it would bring
inflation down to a scale where present or future particle physics
experiments would be able to explore it quite thoroughly. Cosmological
inflation thus enters the realm of testable low energy particle physics.

\section{Conclusions}

We have entered a new era in cosmology, were a host of high precision
measurements are already posing challenges to our understanding of the
universe: the density of ordinary matter and the total amount of energy
in the universe; the microwave background anisotropies on a fine scale
resolution; primordial deuterium abundance from quasar absorption lines;
the acceleration parameter of the universe from high-redshift supernovae
observations; the rate of expansion from gravitational lensing; large
scale structure measurements of the distribution of galaxies and their
evolution, and many more, which already put constraints on the parameter
space of cosmological models [See Fig.~10]. However, these are only the
forerunners of the precision era in cosmology that will dominate the new
millenium, and will make cosmology a phenomenological science.

It is important to bear in mind that all physical theories are
approximations of reality that can fail if pushed too far. Physical
science advances by incorporating earlier theories that are
experimentally supported into larger, more encompassing frameworks. The
standard big bang theory is supported by a wealth of evidence, nobody
really doubts its validity anymore. However, in the last decade it has
been incorporated into the larger picture of cosmological inflation,
which has become the new standard cosmological model. All cosmological
issues are now formulated in the context of the inflationary cosmology.
It is the best explanation we have at the moment for the increasing
set of cosmological observations.

In the next few years we will have an even larger set of high-quality
observations that will test inflation and the cold dark matter paradigm
of structure formation, and determine most of the 12 or more parameters
of the standard cosmological model to a few percent accuracy [See
Table.~1]. It may seem that with such a large number of parameters one
can fit almost anything. However, that is not the case when there is
enough quantity and quality of data. An illustrative example is the
standard model of particle physics, with around 21 parameters and a host
of precise measurements from particle accelerators all over the world.
This model is nowadays rigurously tested, and its parameters measured to
a precision of better than a percent in some cases. It is clear that
high precision measurements will make the standard model of cosmology as
robust as that of particle physics. In fact, it has been the
technological advances of particle physics detectors that are mainly
responsible for the burst of new data coming from cosmological
observations. This is definitely a very healthy field, but there is
still a lot to do. With the advent of better and larger precision
experiments, cosmology is becoming a mature science, where speculation
has given way to phenomenology.

There are still many unanswered fundamental questions in this emerging
picture of cosmology. For instance, we still do not know the nature of
the inflaton field, is it some new fundamental scalar field in the
electroweak symmetry breaking sector, or is it just some effective
description of a more fundamental high energy interaction? Hopefully, in
the near future, experiments in particle physics might give us a clue to
its nature. Inflation had its original inspiration in the Higgs field,
the scalar field supposed to be responsible for the masses of elementary
particles (quarks and leptons) and the breaking of the electroweak
symmetry. Such a field has not been found yet, and its discovery at the
future particle colliders would help understand one of the truly
fundamental problems in physics, the origin of masses. If the
experiments discover something completely new and unexpected, it would
automatically affect inflation at a fundamental level.

One of the most difficult challenges that the new cosmology will have to
face is understanding the origin of the cosmological constant, if indeed
it is confirmed by independent sets of observations. Ever since Einstein
introduced it as a way to counteract gravitational attraction, it has
haunted cosmologists and particle physicists for decades. We still do
not have a mechanism to explain its extraordinarily small value, 120
orders of magnitude below what is predicted by quantum physics. For
several decades there has been the reasonable speculation that this
fundamental problem may be related to the quantisation of gravity.
General relativity is a classical theory of spacetime, and it has proved
particularly difficult to construct a consistent quantum theory of
gravity, since it involves fundamental issues like causality and the
nature of spacetime itself. 

The value of the cosmological constant predicted by quantum physics is
related to our lack of understanding of gravity at the microscopic
level. However, its effect is dominant at the very largest scales of
clusters or superclusters of galaxies, on truly macroscopic scales. This
hints at what is known in quantum theory as an anomaly, a quantum
phenomenon relating both ultraviolet (microscopic) and infrared
(macroscopic) divergences. We can speculate that perhaps general
relativity is not the correct description of gravity on the very largest
scales. In fact, it has been only in the last few billion years that the
observable universe has become large enough that these global effects
could be noticeable. In its infancy, the universe was much smaller than
it is now, and presumably general relativity gave a correct description
of its evolution, as confirmed by the successes of the standard big bang
theory. As it expanded, larger and larger regions were encompassed and
therefore deviations from general relativity would slowly become
important. It may well be that the recent determination of a
cosmological constant from observations of supernovae at high redshifts
is hinting at a fundamental misunderstanding of gravity on the very
large scales.

If this were indeed the case, we should expect that the new generation
of precise cosmological observations will not only affect our
cosmological model of the universe but also a more fundamental
description of nature.

\newpage

\begin{figure}[htbp]
\begin{center}
\hspace*{-1.1cm}
\leavevmode\epsfysize=16cm \epsfbox{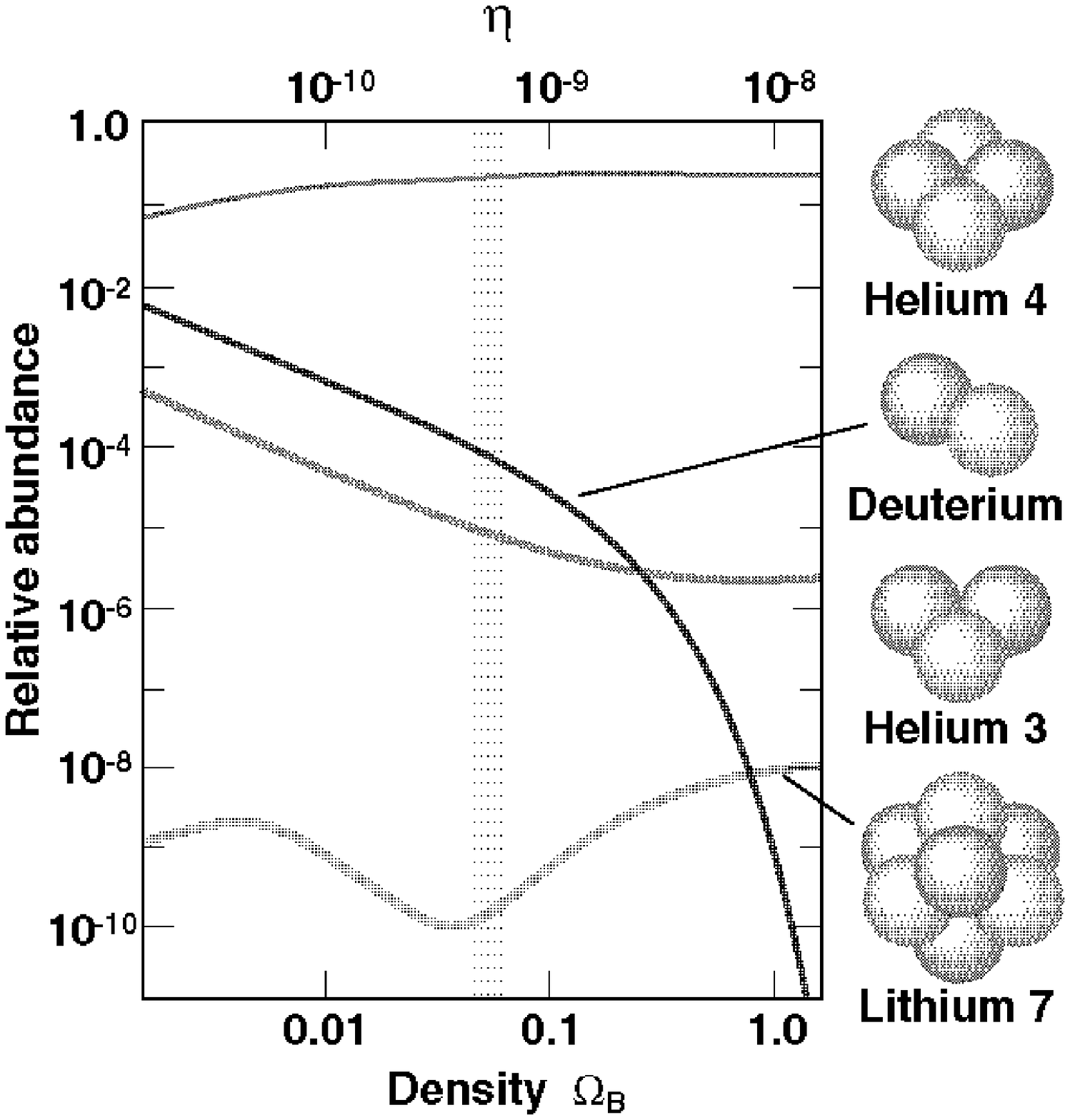}
\vspace*{1cm}
\caption[fig1]{The density of baryons (neutrons and protons) in the
universe determines the relative abundances of the lighter elements,
hydrogen, helium and lithium. For a higher density universe, the
computed helium abundance is little different, but the computed
abundance of deuterium is considerably lower. Thus observations of
primordial deuterium place very strong bounds on the present density of
baryons. The shaded region is consistent with the observations, ranging
over 10 orders of magnitude, from 24 percent for helium to one part in
$10^{10}$ for lithium. This impressive quantitative agreement is one of
the main successes of the standard big bang cosmology. The observed
baryon density $\Omega_B$ corresponds to a baryon to photon ratio $\eta$
of one part in $10^9$.}
\label{fig1}
\end{center}
\end{figure}

\newpage

\begin{figure}[htbp]
\begin{center}
\vspace*{1cm}
\hspace*{-1cm}\leavevmode\epsfysize=15.5cm \epsfbox{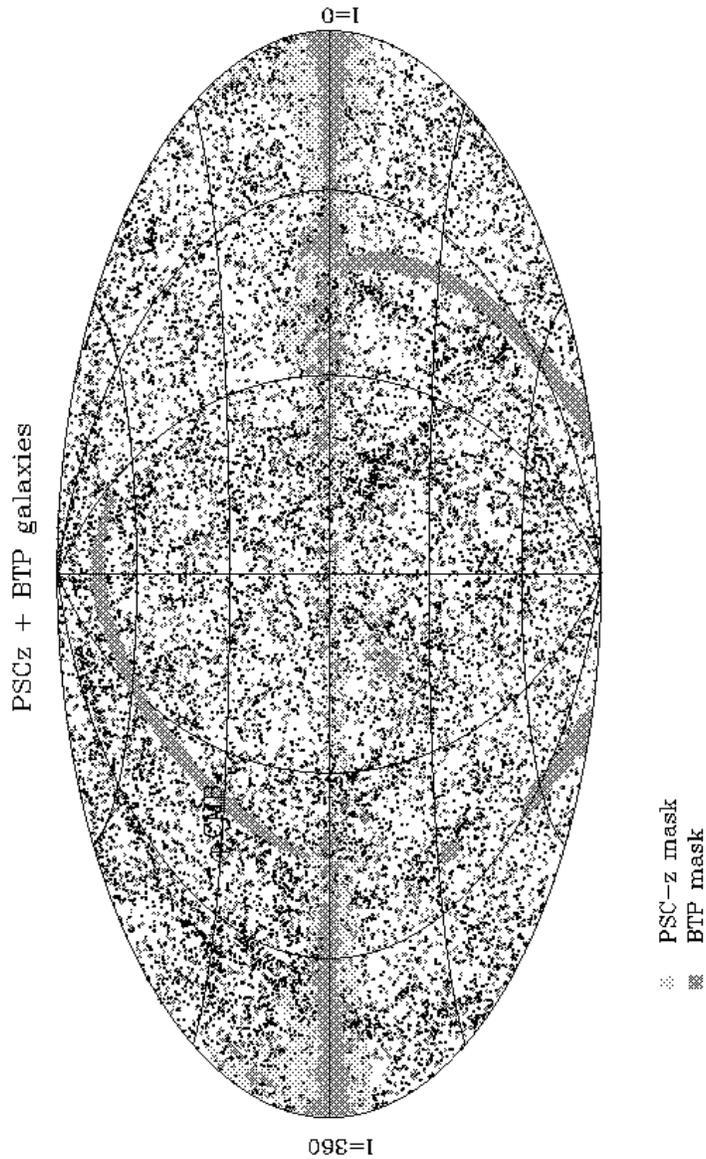}
\vspace*{1.5cm}
\caption[fig2]{The IRAS Point Source Catalog redshift survey contains
some 15\,000 galaxies, covering over 83 percent of the sky up to
redshifts $z\leq0.05$. We show here the projection of the galaxy
distribution in galactic coordinates. The filled-in regions indicate
unobserved or obscured regions, specially along the horizontal strip
surrounding the galactic plane. Future galaxy surveys like the Sloan
Digital Sky Survey will map a million galaxies up to redshifts
$z\leq0.5$.}
\label{fig2}
\end{center}
\end{figure}

\newpage

\begin{figure}[htbp]
\begin{center}
\hspace*{-.1cm}
\leavevmode\epsfysize=10.8cm \epsfbox{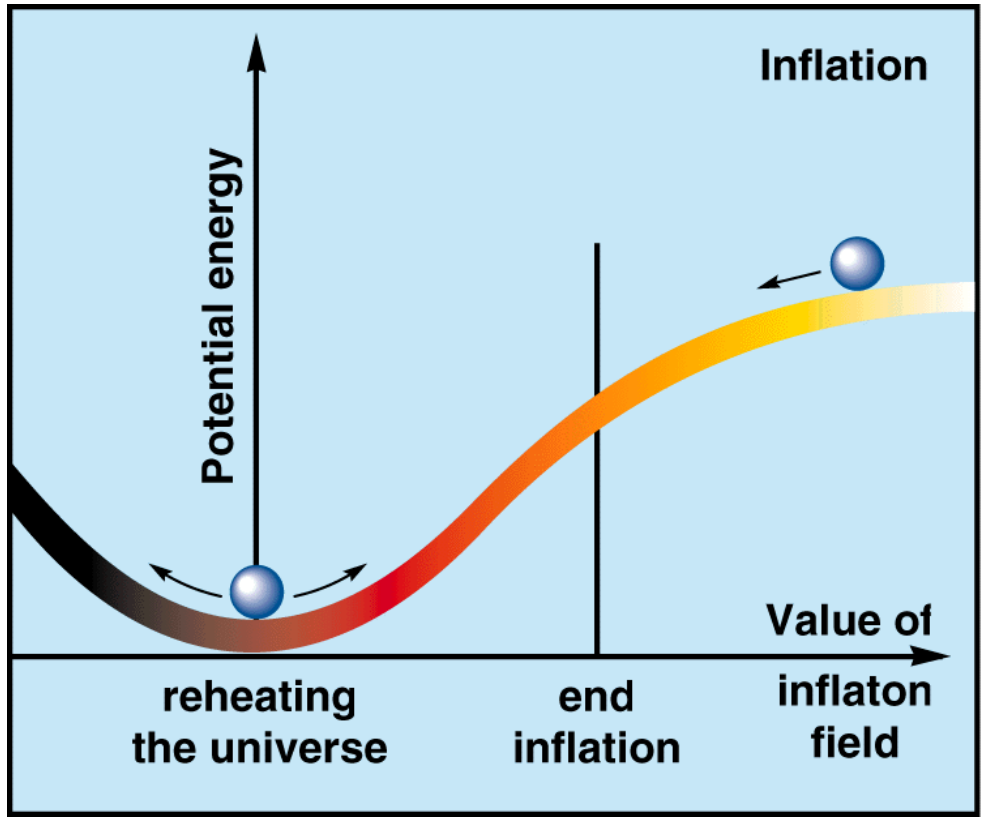}
\vspace*{1cm}
\caption[fig3]{The inflaton field can be represented as a ball rolling
down a hill.  During inflation the energy density is approximately
constant, driving the tremendous expansion of the universe. When the
ball starts to oscillate around the bottom of the hill, inflation ends
and the inflaton energy decays into particles. In certain cases, the
coherent oscillations of the inflaton could generate a resonant
production of particles which soon thermalise, reheating the universe.}
\label{fig3}
\end{center}
\end{figure}

\newpage

\begin{figure}[htbp]
\begin{center}
\vspace*{-4cm}
\hspace*{-1.3cm}
\leavevmode\epsfysize=23cm \epsfbox{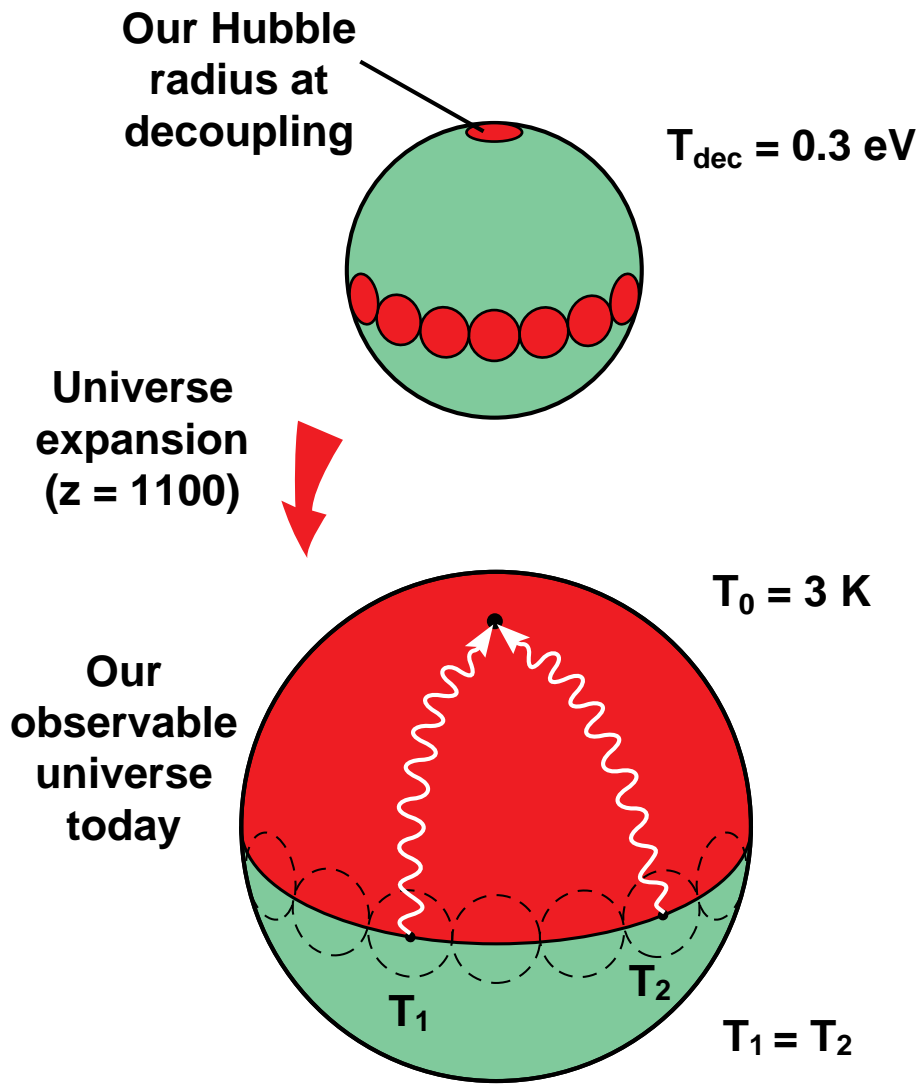}
\vspace*{-2cm}
\caption[fig4]{Perhaps the most acute problem of the big bang model is
explaining the extraordinary homogeneity and isotropy of the microwave
background.  Information cannot travel faster than the speed of light,
so the causal region (so-called horizon or Hubble radius) at the time of
photon decoupling could not be larger than 300\,000 light-years across,
or about one degree projected in the sky today. So why regions that are
separated by more than a degree in the sky should have the same
temperature, when the photons that come from those two distant regions
could not have been in causal contact when they were emitted. This
constitutes the so-called horizon problem, which is spectacularly solved
by inflation.}
\label{fig4}
\end{center}
\end{figure}

\newpage

\begin{figure}[htbp]
\begin{center}
\vspace*{-1cm}
\hspace*{-2.4cm}
\leavevmode\epsfysize=24.2cm \epsfbox{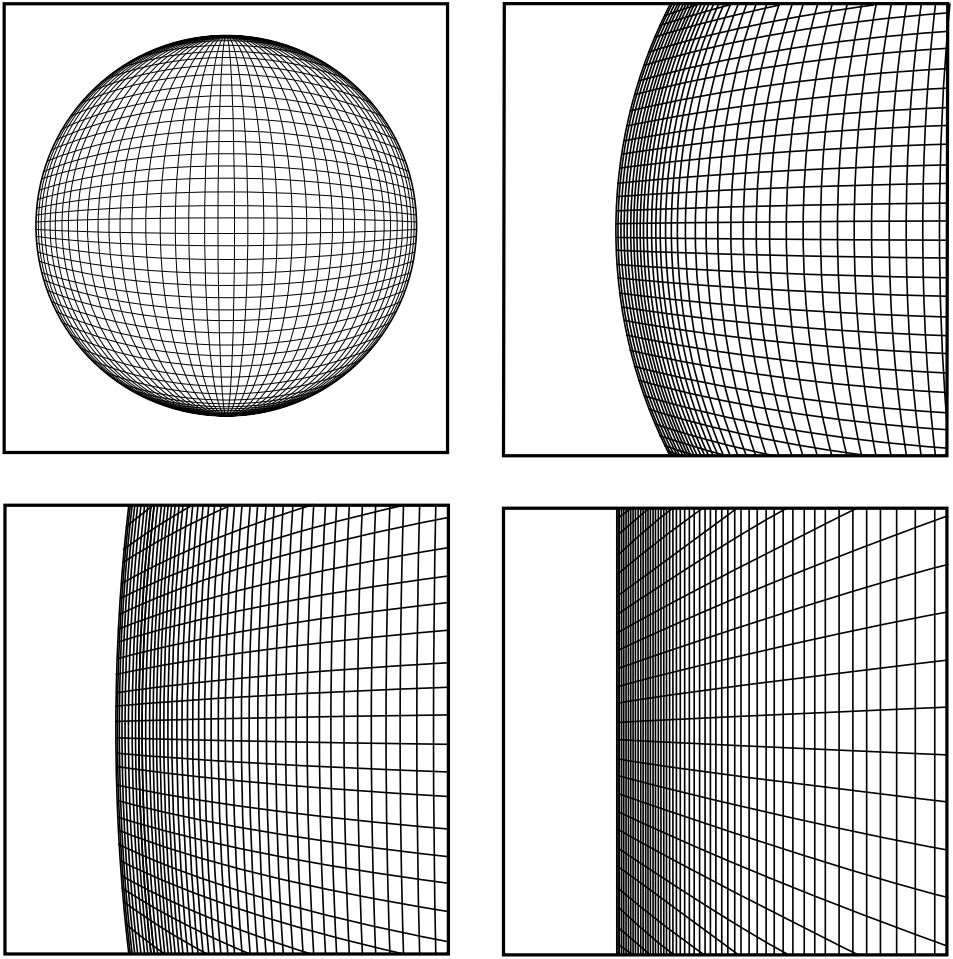}
\vspace*{-8cm}
\caption[fig5]{The exponential expansion during inflation made the
radius of curvature of the universe so large that our observable patch
of the universe today appears essentialy flat, analogous (in three
dimensions) to how the surface of a balloon appears flatter and flatter
as we inflate it to enormous sizes. This is a crucial prediction of
cosmological inflation that will be tested to extraordinary accuracy in
the next few years.}
\label{fig5}
\end{center}
\end{figure}

\newpage

\begin{figure}[htbp]
\begin{center}
\vspace*{-.5cm}
\hspace*{-.5cm}
\leavevmode\epsfysize=17cm \epsfbox{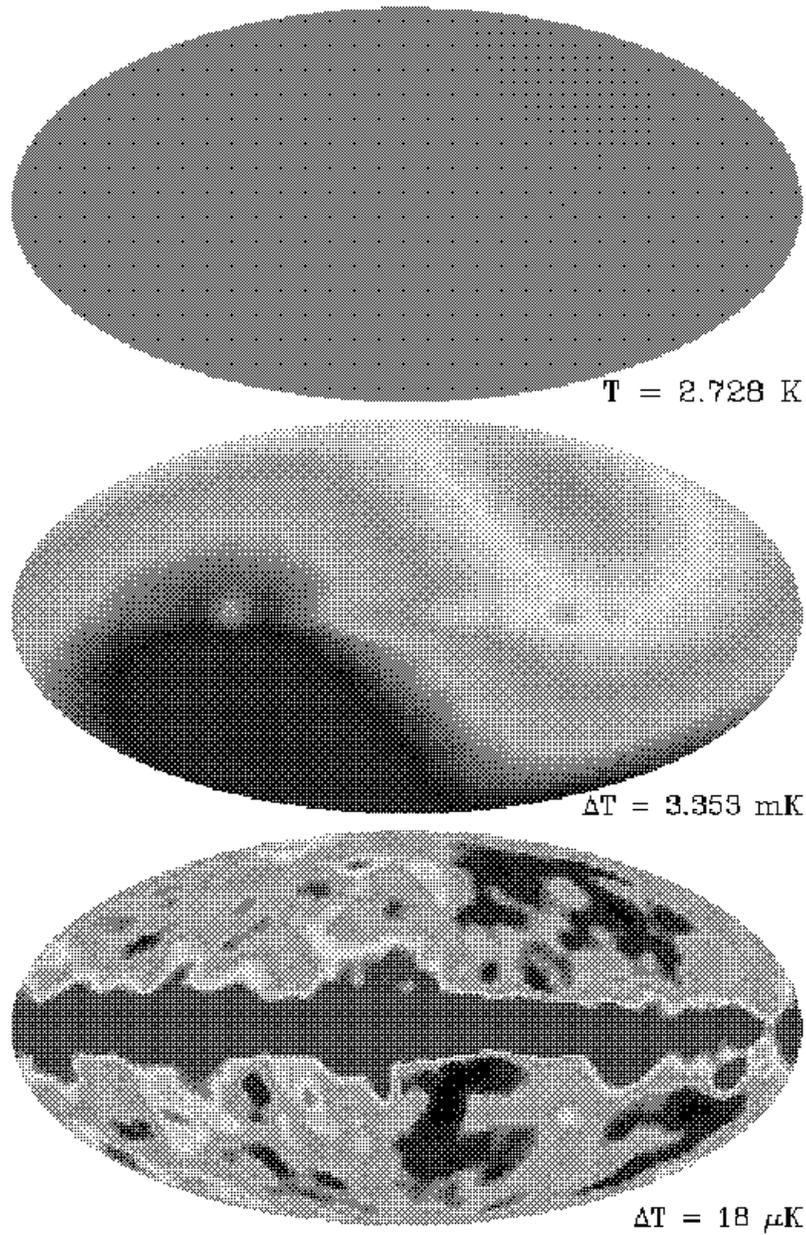}
\caption[fig6]{The microwave background sky as seen by COBE. The upper
panel shows the extraordinary homogeneity and isotropy of the universe:
the microwave background has a uniform blackbody temperature of
$T=2.728$ K. The middle panel shows the dipole, corresponding to our
relative motion with respect to the microwave background in the
direction of the Virgo cluster. The lower panel shows the intrinsic CMB
anisotropies, corresponding to the quadrupole and higher multipoles, at
the level of one part in $10^5$. The horizontal bar (red) corresponds to
the microwave emission of our galaxy, which should be subtracted.}
\label{fig6}
\end{center}
\end{figure}

\newpage

\begin{figure}[htbp]
\begin{center}
\vspace*{.5cm}
\hspace*{-.1cm}
\leavevmode\epsfysize=15.7cm \epsfbox{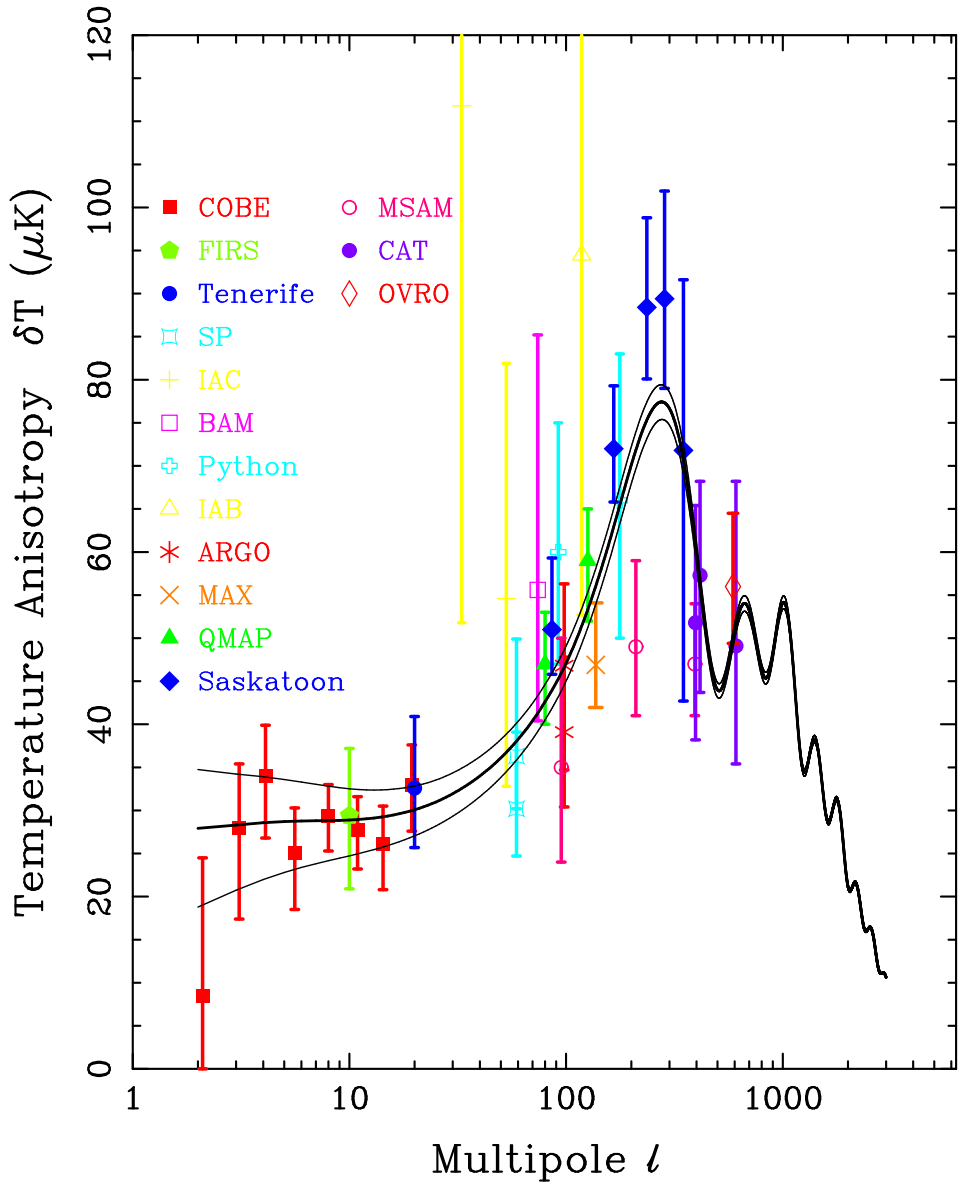}
\vspace*{1cm}
\caption[fig7]{There are at present dozens of ground and balloon-borne
experiments looking at the microwave background temperature anisotropies
with angular resolutions from 10 degrees to a few arc-minutes in the
sky, corresponding to multipole numbers $l=2 - 3000$. Present
observations suggest the existence of a peak in the angular
distribution, as predicted by inflation. The theoretical curve (thick
line) illustrates a particular model which fits the data, together with
the (cosmic variance) uncertainty with which the future satellites MAP
and Planck will be able to measure the microwave background
anisotropies.}
\label{fig7}
\end{center}
\end{figure}

\newpage

\begin{figure}[htbp]
\begin{center}
\vspace*{-3cm}
\hspace*{-1cm}
\leavevmode\epsfysize=17cm \hspace*{-4cm} \epsfbox{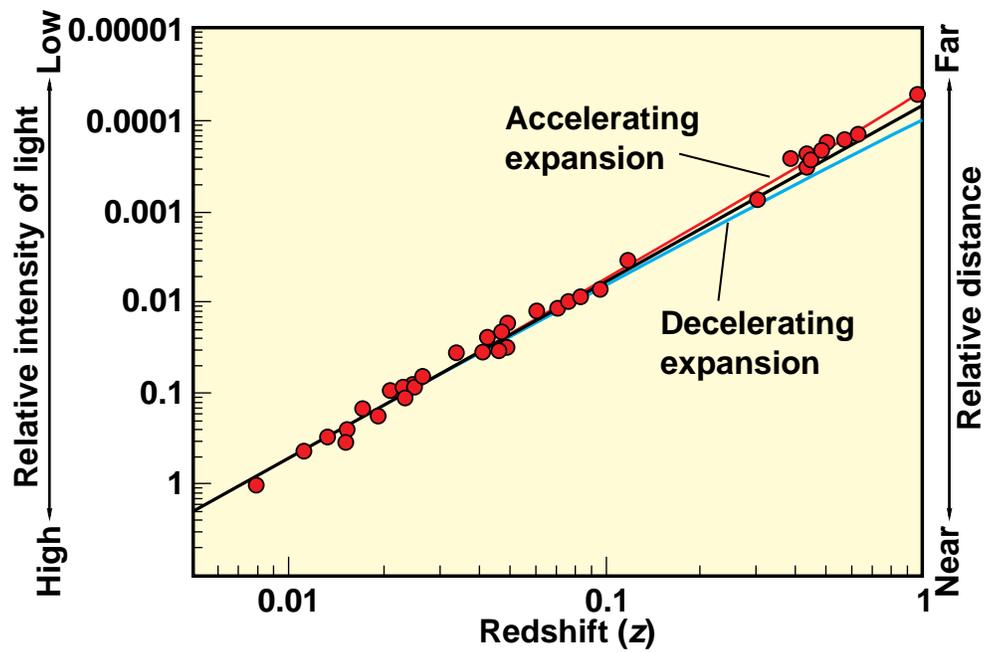}
\vspace*{1.5cm}
\caption[fig8]{The Hubble diagram for high redshift supernovae (red
dots). Observations deviate slightly but significantly from the
Einstein-de Sitter model (blue line), a flat universe with no
cosmological constant, which would be decelerating due to the attraction
of matter. These observations indicate that there is only 30 percent of
the matter necessary to make it flat (black line), and therefore
decelerates more slowly than predicted. The measurements even suggest
that the universe is accelerating (red line), as if due to a nonzero
cosmological constant.}
\label{fig8}
\end{center}
\end{figure}

\newpage

\begin{figure}[htbp]
\begin{center}
\vspace*{-4cm}
\hspace*{-1.6cm}
\leavevmode\epsfysize=23.5cm \epsfbox{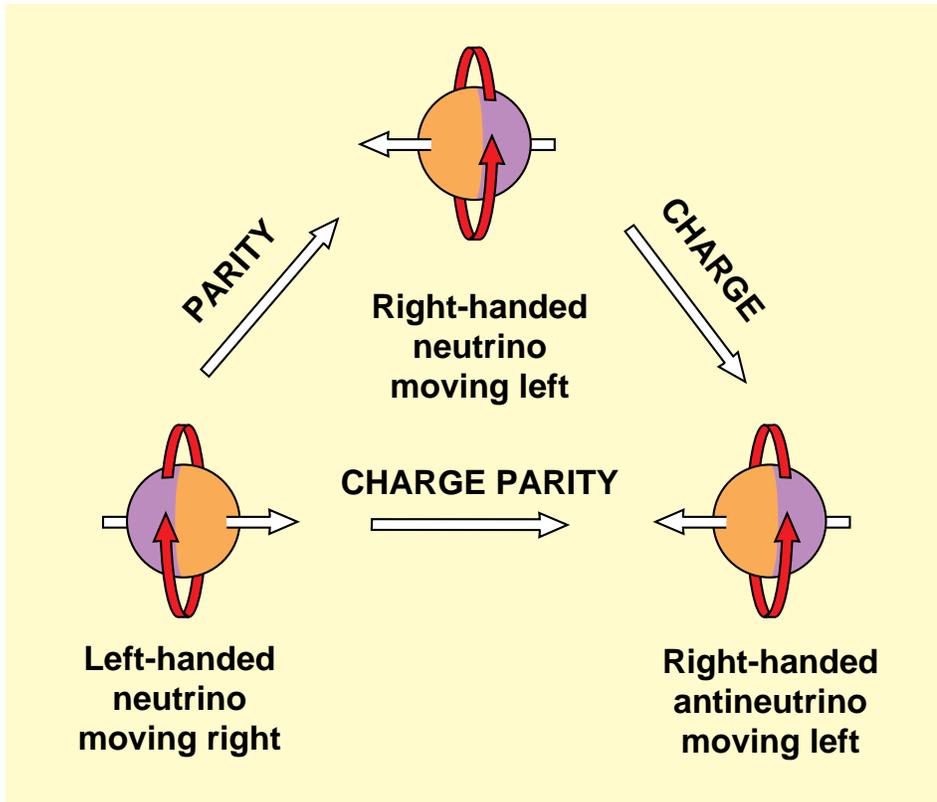}
\vspace*{-6cm}
\caption[fig9]{Symmetries are vital to the study of physics, and few are
more intriguing than the combination of charge and parity. Charge
reversal gives the opposite sign to all quantum numbers such as electric
charge, changing a particle into its antiparticle. Parity reversal
reflects an object and also rotates it by 180 degrees. The laws of
classical mechanics and electromagnetism, as well as the strong
interactions that maintain quarks and nucleons together, are invariant
under either of these symmetry operations. The weak interactions,
however, are changed by both charge and parity reversal. For many years
it appeared that the combination charge-parity together was invariant
even for weak interactions. However, experiments in 1964 shattered this
illusion, posing the puzzled of why nature looks different when
reflected in the charge-parity mirror. It may have an answer in the
origin of the matter-antimatter asymmetry.}
\label{fig9}
\end{center}
\end{figure}

\newpage

\begin{figure}[htbp]
\begin{center}
\vspace*{-6cm}
\hspace*{-2.2cm}
\leavevmode\epsfysize=25cm \epsfbox{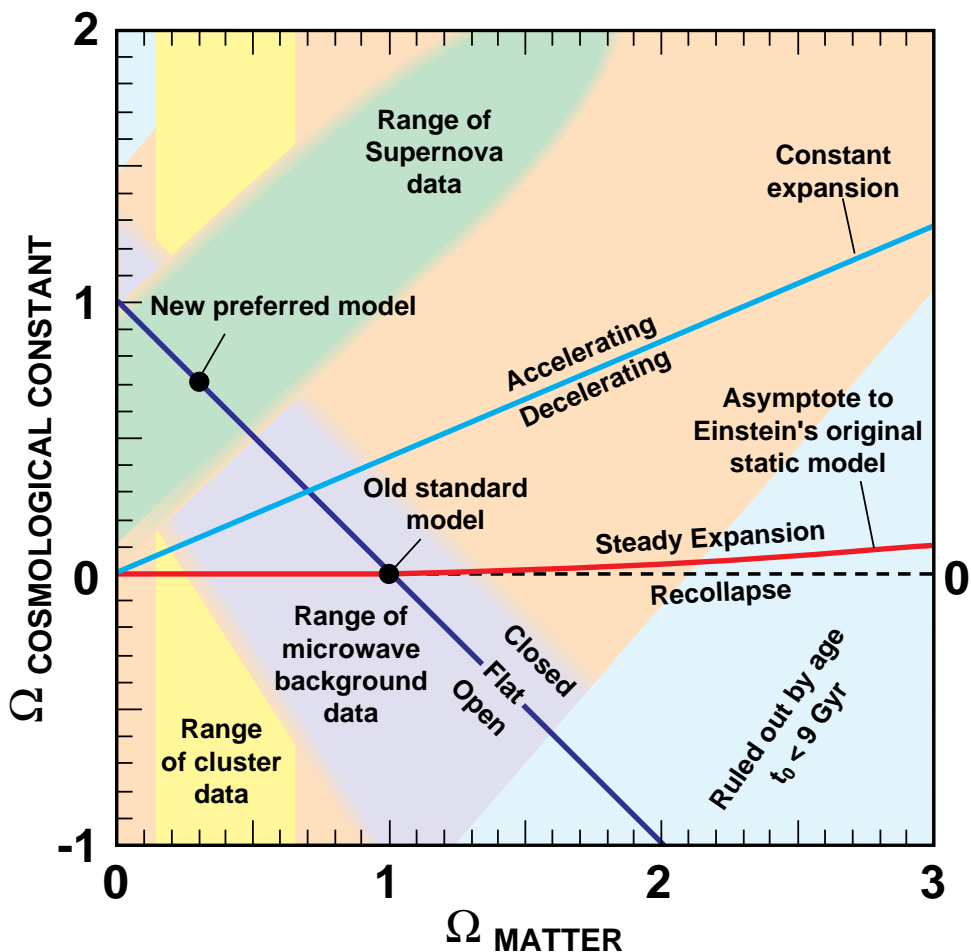}
\vspace*{-3cm}
\caption[fig10]{The evolution of the universe depends on two crucial
parameters: the average density of matter (horizontal axis) and the
energy density in the cosmological constant (vertical axis). Their
values produce three very different effects. First, their sum gives the
total cosmic energy content and determines the geometry of spacetime
(violet line). Second, their difference characterises the relative
strength of expansion and gravity, and determines how the expansion rate
changes with time (blue line). These two effects have been probed by
recent observations (green yellow and purple regions). The third, a
balance between the two densities, determines the fate of the universe
(red line). The three effects have many different
combinations. Surprisingly enough, at present, all observations seem to
lie within a narrow region of parameter space.}
\label{fig10}
\end{center}
\end{figure}

\end{document}